\documentclass[english,fleqn,allpages]{ISTE_science}[2018/07/30]

\usepackage{amsthm}
\usepackage{natbib}

\makeatletter

\newcommand{\bea}{\begin{eqnarray}}
\newcommand{\eea}{\end{eqnarray}}

\newcommand{\cf}{{\it cf.}~}

\newcommand{\ie}{{\it ie.\,}}
\newcommand{\matr}[1]{\mathbf{#1}}
\newcommand{\eps}{\epsilon}

\newcommand{\rw}{\rightarrow}

\usepackage{amsfonts}           
\usepackage{amssymb}            
\usepackage{amsmath}            
\usepackage{comment}
\usepackage{latexsym}           
\usepackage{dcolumn}            
\usepackage{epstopdf}           
\usepackage{multirow}
\usepackage{adjustbox}
\usepackage{url}
\usepackage{MnSymbol}
\usepackage{subfigure}

\newsavebox{\fminibox} \newlength{\fminilength}

\title{Effects of Rate, Size and Prior Deformation in Microcrystal Plasticity}
\begin{document}

\raggedbottom
\mainmatter
\chapter{Effects of Rate, Size and Prior Deformation in Microcrystal Plasticity}
\label{chap-struct}

\markboth{Effects of Rate, Size and Prior Deformation in Microcrystal Plasticity}{Effects of Rate, Size and Prior Deformation in Microcrystal Plasticity}

\authorname{Stefanos \Name{Papanikolaou}, Michail \Name{Tzimas}}{West Virginia University}

\section{Introduction}

\label{sec:Introduction}

In macroscale mechanics, it is natural to define constitutive laws for the mechanical response of a particular material class and/or geometry. A characteristically simple example is the deformation of a crystalline sample, described by Hooke's law for the elastic regime and a yield surface with a plastic flow rule for the inelastic one. {   In the extreme limit of small finite volumes, while Hooke's law persists, the concepts of a yield surface and smooth plastic flow are controversial, manifesting into mechanical properties' strong dependence on size, rate, and prior deformation. }.

Characterizing the extent of failure of traditional inelastic constitutive laws, in relation to phenomena at sub-micron length scales, has been a consistent focus of material science over the last two decades. The key aspect has been the { understanding of the effects of strain gradients, intrinsic or not~\citep{hutchinson2000plasticity}, on  plastic deformation of crystals in various geometries, the most prominent of which has been nanoindentation~\citep{oliver_pharr_2010}. In small finite volumes~\citep{Uchic03}, the focus mainly has been  the investigation of uniaxial compression in micro and nano-pillars~\citep{uchic2009plasticity,Greer:2011aa}: Size dependence has been evident in the material strength due to intrinsic defect-induced strain gradients~\citep{proceedingsNix}, while rate dependence has been strongly suspected due to the fact that plastic response displays strong intermittent features~\citep{Papanikolaou2012}, {   typically labeled as avalanches~\citep{Zaiser2006}. Moreover, the well accepted phenomenon of ``mechanical annealing"~\citep{shan2008mechanical}, namely the drastic increase of a pillar's strength through prior compression, has unraveled a well suspected but elusive, strong connection between small finite volume's initial conditions and prior deformation history of micro-sized specimens~\citep{CRAT:CRAT2170190611}. }

For the theoretical investigation and explanation of crystal plasticity in the uniaxial compression of  nano-sized specimens, modeling efforts have spanned the whole multiscale modeling spectrum. The investigation of the combined effects of all possible material and geometry details led to atomistic and molecular simulations~\citep{yamakov2004deformation,Rabkin:2007eb} that are limited at ultra-high strain-rates and tiny loading volumes. These studies have unveiled various delicate features of relevant dislocation mechanisms, such as surface dislocation multiplication. For collective dislocation behaviors, three-dimensional dislocation dynamics simulations have been utilized for the relevant mechanisms behind observed size effects~\citep{Greer:2011aa,kraft2010plasticity,uchic2009plasticity}, rate effects~\citep{maass2017micro, Papanikolaou2012} and also various statistical aspects such as avalanche size distributions~\citep{Cui2016}. However, the demanding nature of the simulation of realistic micropillar dislocation densities ($10^{14}$/m$^2$~\citep{shan2008mechanical}) in sub-micron volumes ($<5\mu$m) at experimentally relevant strains ($\sim 5\%$) has led to limited statistical sampling of initial conditions with limited dislocation line topologies. An assisting role to all these methods and results,  has been promoted for two-dimensional discrete dislocation simulations, which have been providing reliable statistical features and insightful connections, with available, benchmarked, theoretical constructions~\citep{Zaiser:2015ri,Groma:2016pz} and significant experimental validation~\citep{nicola2006plastic}. In this context, the use of periodic boundary conditions has allowed a remarkable level of statistical fidelity ~\citep{alava2014crackling} and agreement with other models of statistical mechanics~\citep{Zaiser2006,Yefimov:2004qz,ovaska2015quenched}. However, such 2D-DDD models either function at low dislocation densities or compromise the physical constraint of finite-volume boundary conditions. 

In the pursuit of a self-consistent explanation of experimental phenomenology that  maintains the constraint of finite volume boundary conditions, as well as reliable statistical sampling, a minimal 2D-DDD model has been proposed, benchmarked to experimental findings~\citep{Papanikolaou2017}, and studied for aspects of rate~\citep{PhysRevLett.122.178001}, size~\citep{Papanikolaou2017} and initial-condition/prior-deformation dependence~\citep{pap19}. The discussion of the findings in this 2D-DDD model is the focus of this chapter. While 2D in character, this model achieves to capture the statistical behavior of micron-sized specimens, using appropriate finite-volume boundary conditions. The initial conditions are thoroughly investigated by consistently generating statistically averaged pre-existing dislocation microstructures due to prior compressive/tensile deformation. It is worth noting that such statistically averaged  finite-volume dislocation microstructures are {\it unique} in dislocation simulations across scales and simulation methods. Besides explaining existing experimental findings in the context of rate, size and statistical averaging aspects, this model has led to significant predictions for the dislocation density dependence of size effects, the character of rate effects, the fractal features of the specimen boundaries, as well as the development of experimentally relevant machine learning methods that may apply to more general problems in mechanics of materials.

\section{Model}
\label{sec:Model}

We consider a \emph{minimal} model of crystal plasticity~\citep{Papanikolaou2017} for uniaxial compression of $Al$ thin samples (Young's Modulus $E = 70$ GPa, Poisson ratio $\nu = 0.33$ with the equivalent Young's Modulus for plane strain problems being~$E^*~=~E/(1-\nu^2)~=~78.55$~GPa and the Burger's vector length being $b = 0.25$nm), which captures the energetics of crystal deformation mediated through gliding of edge dislocations along one or multiple slip systems. The dislocation mobility parameter $B$ is set to $B= 10^{-4} Pa\cdot s$. In the model, gliding of dislocations occurs in slip planes separated by $10b$, oriented at $\pm 30^\circ$ from the loading direction (Fig.~\ref{fig:schematic} (a). 

Slip planes may become active only when they contain at least one source for dislocation generation. Bulk sources are randomly but uniformly distributed over slip planes, and their strength is selected randomly from a Gaussian distribution with mean value $\tau_{nuc} = 50$ MPa and 10 $\%$ standard deviation. {   Sources are randomly distributed with density $\rho^{\rm{bulk}}_{\rm{nuc}}= 60 \mu m^{-2}$.
Dislocations are generated from sources when the resolved shear stress $\tau$ at the source location is sufficiently high ($\tau>\tau_{\rm nuc}$) for a sufficiently long time $t_{\rm nuc}$. The model considers only gliding of dislocations, so the dislocation motion is solely controlled by the component of the Peach-Koehler force in the slip direction. 

Point obstacles are randomly distributed on active slip planes with a constant density that corresponds on average, 8 randomly-distributed obstacles per each bulk dislocation source (\ie $\rho^{\rm{bulk}}_{\rm{obs}}= 480 \mu m^{-2}$). In this way, the source and obstacle densities remain statistically similar as finite volume dimensions change.
Obstacles account for precipitates and forest dislocations on out-of-plane slip systems. Our simple obstacle model is that a dislocation stays effectively pinned until its Peach-Koehler force exceeds the obstacle-dependent value $\tau_{\rm obs}b$. The strength of the obstacles $\tau_{\rm obs}$ is taken to be $300$ MPa with 20\% standard deviation, to account for large variability in realistic scenarios of dislocation pinning.

A model volume may be seen in Fig.~\ref{fig:schematic}, where slip planes (lines) span the sample, equally spaced at $d=10b$. Planes close to corners are deactivated to maintain a smooth loading boundary. Initially, samples are stress free and mobile-dislocation free, and the aspect ratio of height $h$ over width $w$ is maintained constant for all samples, $a=h/w=4$. Dislocations can either exit the sample through the traction-free sides, annihilate with a dislocation of opposite sign when their mutual distance is less than $6b$, or become effectively pinned at an obstacle. 
}
The simulation is carried out incrementally, using a time step that is a factor 20 smaller than the nucleation time $t_{\rm nuc}=10$~ns. At the beginning of every time increment, nucleation, annihilation, pinning at and release from obstacle sites are evaluated. After updating the dislocation structure, the new stress field in the sample is determined, using the finite element method to solve for image fields~\citep{vandergiessen1995}.

\begin{figure}[!ht] \centering
	\includegraphics[width=\textwidth]{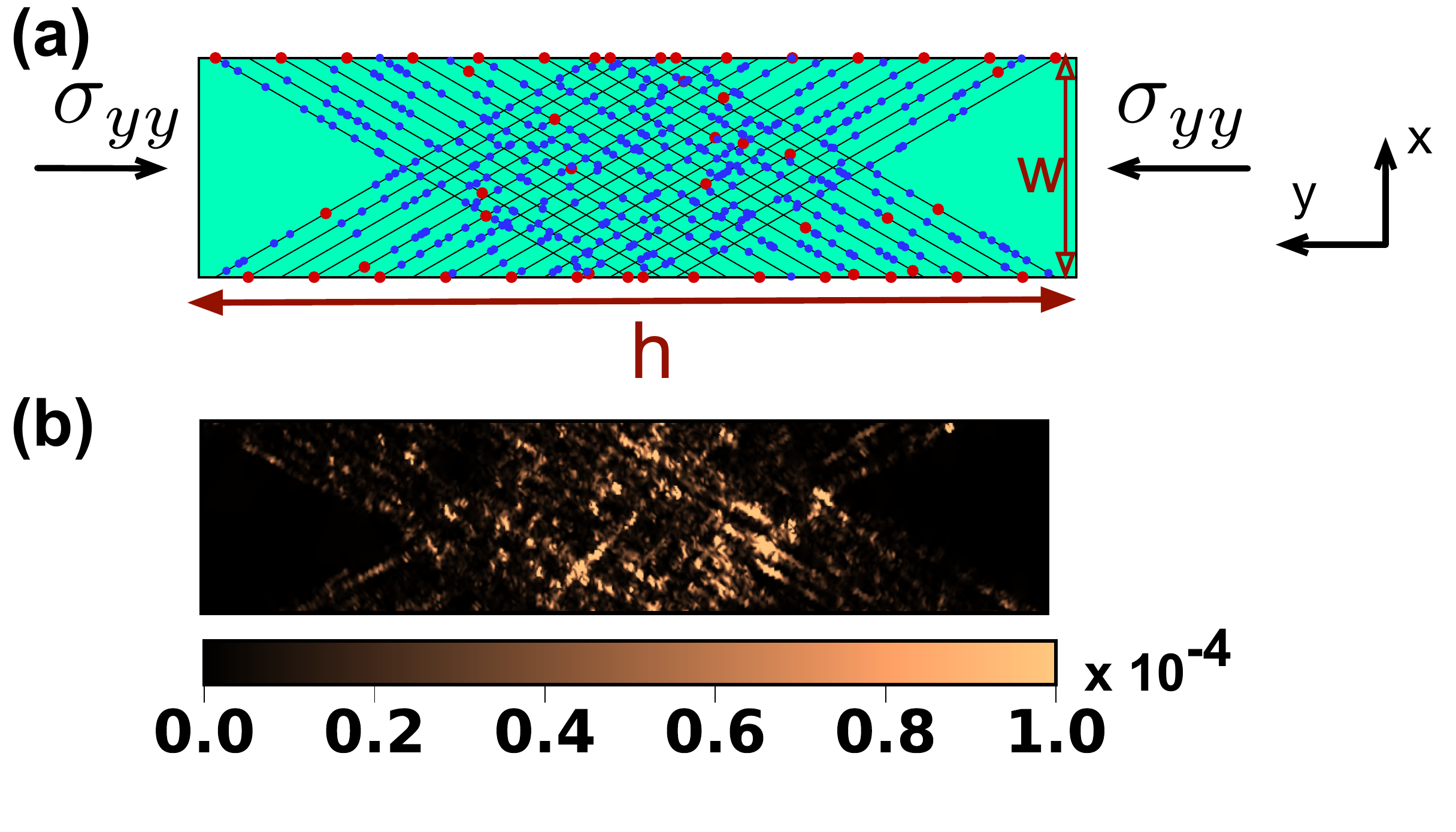}
	\caption{\textbf{Simulation of uniaxial compression of thin films:} (a) The 2D discrete dislocation plasticity model of uniaxial compression of thin films. Slip planes (lines), surface and bulk dislocation sources (red dots) and forest obstacles (blue dots) are seen. (b) Strain profile of sample of $w = 2 \ \mu m$, upon reloading at 0.1\% strain. Initial loading at 10\% strain.}
	\label{fig:schematic}
\end{figure}

{   Overall, the model is based on the singular theory of dislocations, but dislocations may never overlap into dislocation junctions, instead they follow the rules presented for dislocation annihilation.  Similarly, dislocation nucleation is performed at a lengthscale where the dislocation dipole is stable, and the complexity induced by dislocation singulatirites disappear. In this way, no singularities are never encountered during simulation. A phenomenological comparison to expreriments using single crystals was performed in~\citep{Papanikolaou2017}, and displays qualitative agreement that involves not only strengthening effects but also noise observations. No comparisons to 2.5 DDD simulations have been performed. However, it is worth noting that the model agrees in various ways with previous 2D simulations which were compared to experiments such as~\citep{nicola2006plastic}.}
\section{Effects of loading rates and protocols in crystal plasticity}
\label{sec:loading_modes}

The differences between strain-controlled loading (SC) and displacement-controlled loading (DC) rates have been known to be absent at small loading rates in crystal plasticity~\citep{asaro2006mechanics}. However, in small finite volumes, due to the very existence of abrupt avalanche phenomena, there has been evidence and suspicion~\citep{Papanikolaou2012} for significant but {\it statistical} rate dependent effects. The detailed rate effects that originate in the distinct loading protocols have been studied recently for uniaxial compression of micropillars~\citep{maass2015slip, sparks2018shapes} where several rate-dependent scaling behaviors were identified for rates higher than $10^2$/s. 

At the macro-scale, crystals are known to display strong rate effects due to viscoplastic {\it dislocation drag effects} as the strain rate surpasses $\sim 5000/s$~\citep{armstrong2008high,murphy2010strength,tong1992pressure,clifton2000response}. This increase in flow stress has also been seen in DDD simulations~\citep{agnihotri2015rate, hu2017strain} and stems from a natural competition between two timescales in dislocation dynamics. {   The first timescale refers to the dissipative motion of a dislocation inside the crystal (dislocation drag). The second timescale refers to the dislocation nucleation process from a randomly placed source~\citep{hirth1982theory}. Nucleation of dislocations is particularly important for small-scale plasticity. 

These two timescales minimally represent two natural and distinct possibilities in the complex landscape of possible dislocation processes. }The competition of these two timescales should extend in small finite volumes, providing a transition regime around $10^3$/s loading rates, thus a statistically reliable study in loading strain-rates $\dot{\epsilon}$ from $10$/s to $10^{5}$/s would suffice~\citep{PhysRevLett.122.178001}. In the case of pure elasticity, SC and DC loading modes can be compared by using $\dot{\sigma}=E^{*}\dot{\epsilon}$, where $\dot{\sigma}$ is the stress rate and $\dot{\epsilon}$ is the strain rate. Typical simulation parameters are listed in Table~\ref{tab:i}.

\begin{table}[!ht]\Large \centering
	\caption{{{\bf Model parameters for the study of rate effects in uniaxial compression}: Slip plane spacing $d$, slip plane orientation $\theta$, source density $\rho_{\rm{nuc}}$, average source strength $\bar{\tau}_{\rm{nuc}}$, nucleation time $t_{\rm{nuc}}$, obstacle density $\rho_{\rm{obs}}$, average obstacle strength $\bar{\tau}_{\rm{obs}}$.}~\citep{PhysRevLett.122.178001}. }
	\begin{adjustbox}{max width=\textwidth}
		\begin{tabular}{| c | c | c |}
			\hline
			slip planes &sources &  obstacles  \\
			\hline
			$d$=10$b$ & $\rho_{\rm{nuc}}$=60$\mu m^{-2}$ & $\rho_{\rm{obs}}$=480MPa  \\
			$\theta=30^\circ$ & $\bar{\tau}_{\rm{nuc}}=$50MPa  & $\bar{\tau}_{\rm{obs}}=$150MPa \\
			& $\delta\tau_{\rm nuc}=$5MPa  & $\delta\tau_{obs}=$20MPa \\
			\hline
		\end{tabular}
	\end{adjustbox}
	\label{tab:i}
\end{table}

Timescale competitions are generic in most non-equilibrium systems~\citep{sahni1983kinetics} and one may devise simple non-linear dynamical models to explain the basic effects. For example, one may consider a minimal model for the strain evolution due to a dislocation segment that may or may not be trapped into a dislocation source, $d\eps/dt = \sigma + \mu \eps - \eps^3$, where $\eps, \sigma$ are scalars resembling strain and stress variables, and $\mu$ is a mobility parameter. The mobility parameter should have a different sign dependent on the dislocation trapping status. In the absence of dislocation interactions, on a slip plane with a single mobile dislocation, the mobility parameter is $\mu=\mu_{\rm drift} < 0$, and the time for stress $\sigma$ relaxation inside the volume in every incremental timestep is  $\delta t_{\rm drift}=|\mu_{\rm drift}|^{-1}$. In contrast, if there exists a dislocation source but not any mobile dislocations on the slip plane, then the mobility parameter becomes  $\mu=\mu_{\rm nuc}>0$ and the corresponding timescale is $\delta t_{\rm nuc}=\mu_{\rm nuc}^{-1}$. In most cases, the association between these timescales is $\delta t_{\rm nuc}\gg\delta t_{\rm drift}$, so stress increments are accommodated by nucleation events. However, if a system contains multiple dislocation sources, dislocation interactions may frustrate the system due to the disparity of relaxation time and cause a complexity in the evolution dynamics.
In the aforementioned model~\citep{PhysRevLett.122.178001}, dislocations have mobility $\mu_d$ driven by local stress-induced forces~\citep{hirth1982theory}. Gliding of dislocation occurs in a single slip system (slip planes oriented at $30^{\circ}$, see Fig.~\ref{fig:1} (a)). In Figure~\ref{fig:1} (b) stress-strain curves of SC can be seen for low, $10^2$ (blue line) and high $10^5$ (green line) stress rates. Correspondingly, the strain patterns at the same final strain (5\%) are seen in Fig.~\ref{fig:1}~(c),~(d) where the plasticity is localized at low stress rates and is uniform at higher loading rates. 
\begin{figure}[t] \centering
	\includegraphics[width=\textwidth]{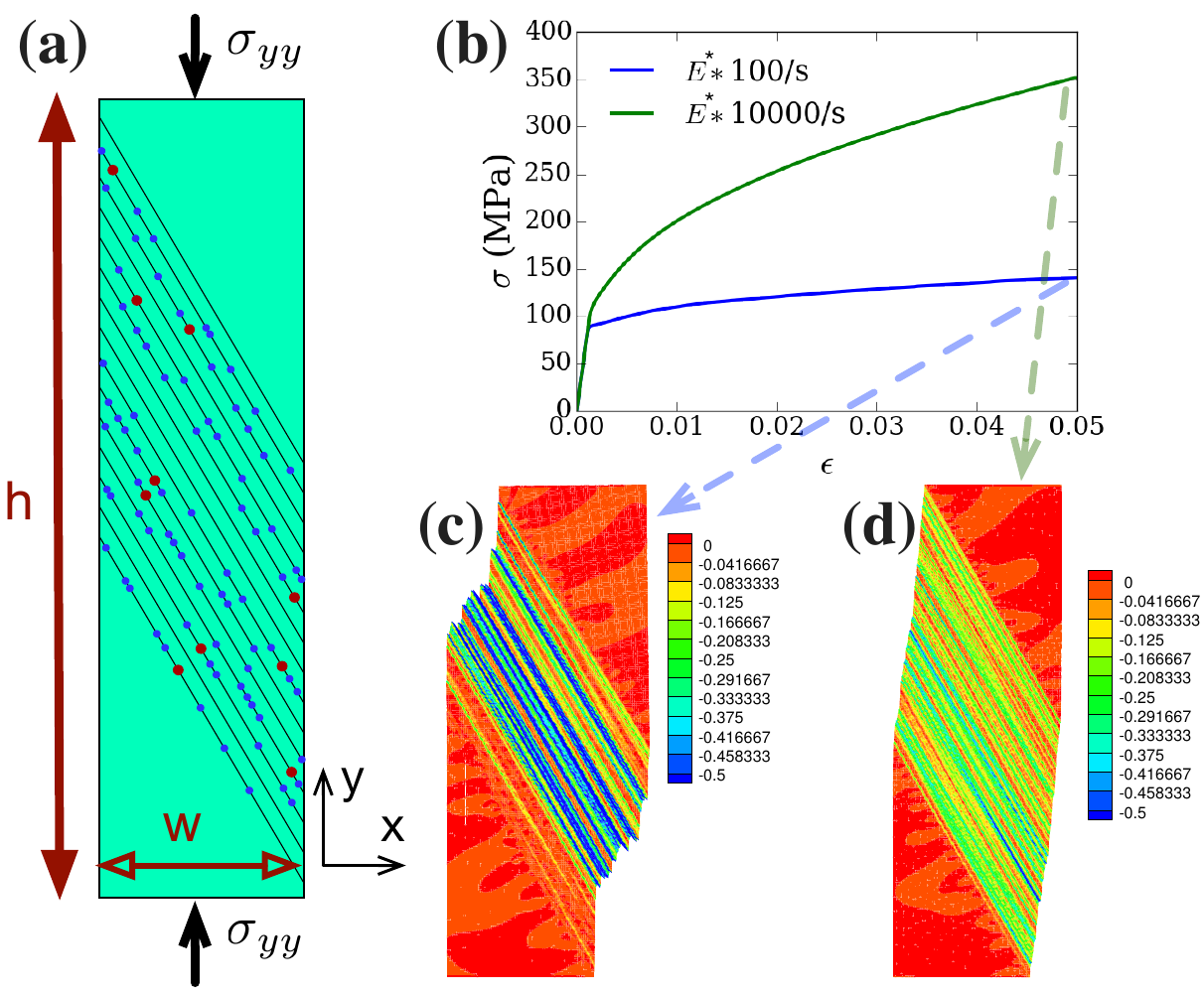}
	\caption{{\bf Rate effects on thin films. }{\bf (a)} The pillar under compression (single slip system). {\bf (b)} Sample stress strain curves of compression at high ($10^5/$s) and low ($10^2/$s) stress rates $\dot{\sigma}$. {\bf (c)} Stain pattern for low $\dot{\sigma}$,  {\bf (d)} Strain pattern for high $\dot{\sigma}$.~\citep{PhysRevLett.122.178001}}
	\label{fig:1}
\end{figure}
\begin{figure}[t] \centering
	\includegraphics[width=\textwidth]{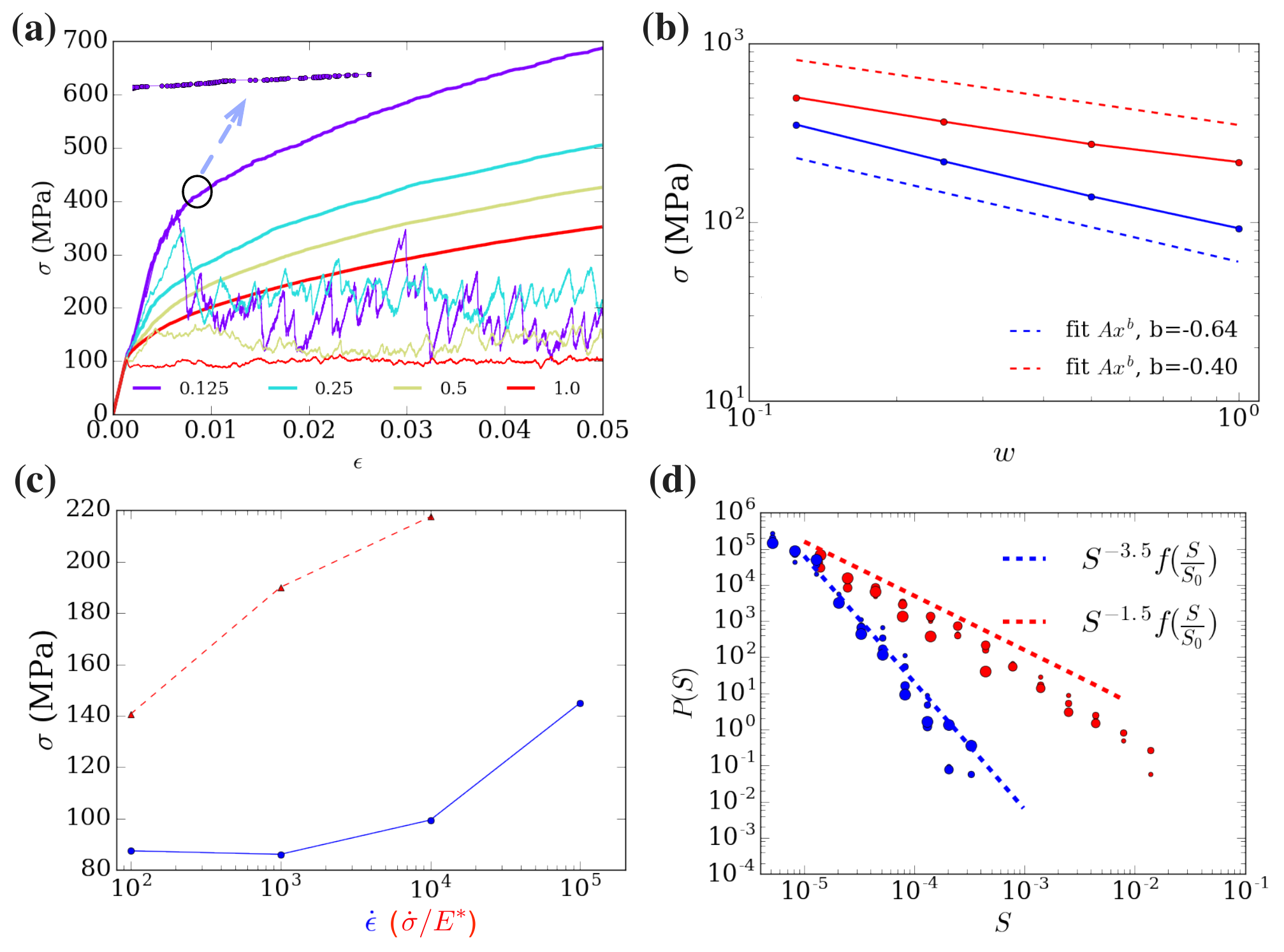}
	\caption{ {\bf Effect of loading protocol: Stress-Controlled (SC) vs. Displacement-Controlled (DC).} Blue curves are DC and red curves SC.
		{\bf (a)} Stress-strain curves of different $w$ using two different loading protocols.Strain bursts are shown; {\bf (b)} Size effect of flow stress at 2\% strain. {\bf (c)} Dependence of flow stress (for $w=1\mu\rm{m}$) on rate. {\bf (d)}: Events (strain jumps) statistics for different loading protocols
		~\citep{PhysRevLett.122.178001}}
	\label{fig:2}
\end{figure}
In Figure~\ref{fig:2}~(a), $\dot{\epsilon} = \ 10^4$/s in DC and correspondingly $\dot{\sigma} = \ E^{*}*10^4$/s.  One may notice the onset of expected work hardening in SC conditions, while in DC conditions one observes softening, with the difference becoming more pronounced as the system width decreases. The model also displays consistent size effects~\citep{papanikolaou2017avalanches,Papanikolaou2017} ($\sigma_Y\sim w^{-0.4-0.6}$)  for both loading protocols~(\cf Fig.~\ref{fig:2} (b)) for average flow stresses (at 0.2$\%$ engineering strain) of 50 realizations. 

Fig.~\ref{fig:2} (c) shows that a flow stress rate dependence is observed in both DC and SC loading modes, even though DC shows a weaker dependence. Upon closer examination of Fig.~\ref{fig:2} (c)~\citep{PhysRevLett.122.178001} one finds that low SC rates {\it statistically resemble} larger DC rates. The origin of this strain-rate crossover is hidden in the amount of strain that nucleation events can accommodate, with $\dot\eps>10^{3}/s$ forcing dislocation drag to take over in the dynamics of dislocations instead of dislocation nucleation. This is consistent with metallurgy phenomenology~~\citep{follansbee1988constitutive,tong1992pressure,clifton1990high}.  While both DC and SC display a flow stress rate effect, their statistical noise behavior is very different; evidence arises from the study of the (SC) strain jump statistics in Fig.~\ref{fig:2} (d): In SC, event size is defined as $S=\sum_{i\; \in\; \left\lbrace\delta\epsilon^{i}>\epsilon_{\rm{threshold}}\right\rbrace}\delta \epsilon^i$; In contrast, in DC, an event is characterized by stress drops $\delta\sigma$ which lead to temporary displacement overshoots -- thus, in order to compare the two loading conditions, a DC strain burst event size is defined as $S=\sum_{i\; \in\;\left\lbrace {-\delta \sigma^{i}}>\sigma_{\rm{threshold}} \right\rbrace}\delta \epsilon^i$~\citep{Cui2016}.

\begin{figure}[t] \centering
	\includegraphics[width=\textwidth]{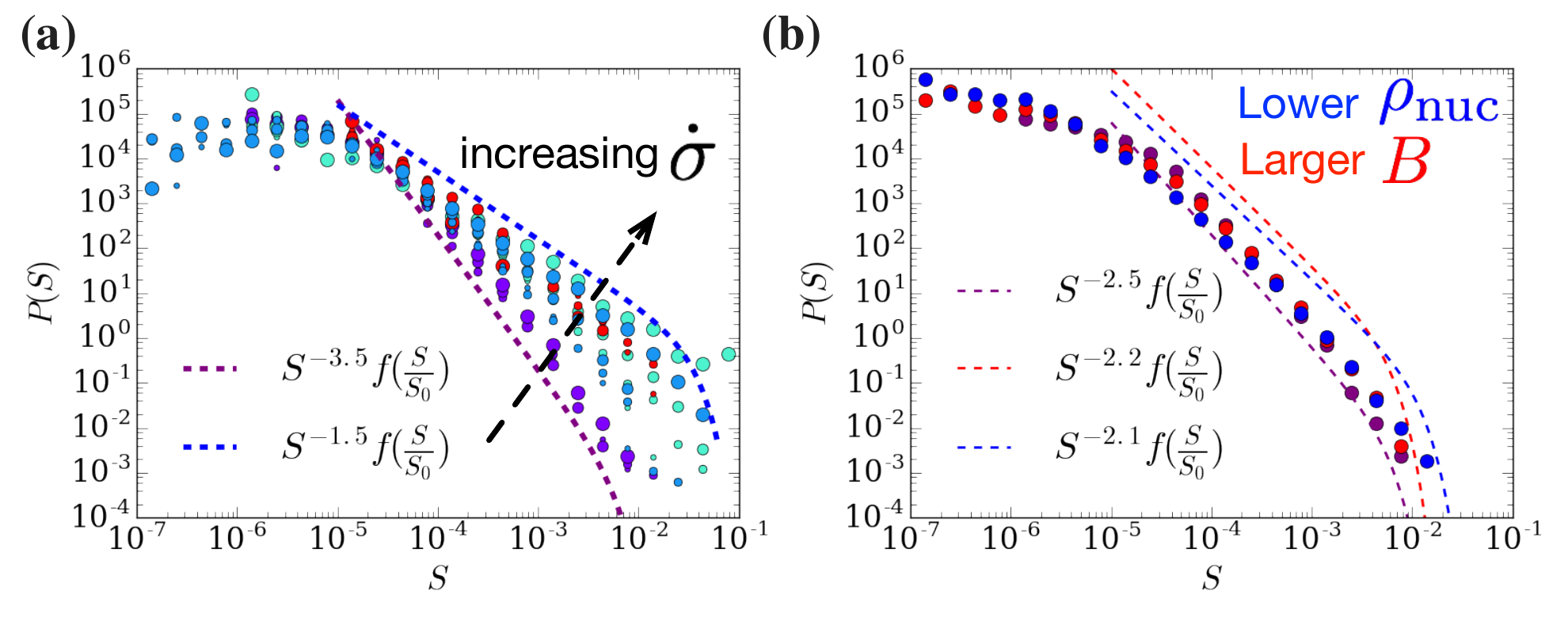}
	\caption{{\bf SC Rate Effect Crossover.~}
	{\bf(a):} Event statistics for different $\dot{\sigma}$ using SC. {\bf (b):} Effect of dislocation source density $\rho_{\rm{nuc}}$ and mobility $B$ on power law exponent.}
	\label{fig:3}
\end{figure}

The model has two intrinsic time scales~\citep{agnihotri2015rate}: the dislocation nucleation timescale $\delta t_{\rm{nuc}} = \ 10~ns$, which can be associated to the dislocation multiplication timescale in other models of plasticity, and the ``drag" timescale which may be defined ia the ratio between dislocation mobility and material Young's modulus $B/E$. In this model, the {\it drag timescale} is  $10^{-6}$ ns, consistent with single-crystal thin film experiments for the moduli and dislocation mobility~\citep{xiang2006bauschinger,nicola2006plastic}. As shown in  Fig.~\ref{fig:2} (d), plastic events' statistics can be estimated through the analysis of the stress strain curves shown in Fig.~\ref{fig:2} (a); histograms of sizes have different $\tau$ exponents with consistent power law behavior: $\tau$ is close to $3.5$ for DC and 1.5 for SC. Another interesting fact is that this exponent difference decreases as the stress loading rate increases: In Figure~\ref{fig:3} (a) we see statistics for different stress rates varying from $\dot{\sigma}=E^{*}*10$/s to $\dot{\sigma}=E^{*}*10^{4}$/s. Power law events distribution appear for all stress rates, yet with different exponent which changes from 3.5 for $\dot{\sigma}=E^{*}*10$/s to 1.5 for $\dot{\sigma}=E^{*}*10^{4}$/s. This  dependence of exponents on the stress rate indicates a non-trivial connection between the event statistics and the transition from {\it nucleation-dominated} to {\it drag-dominated} dislocation dynamics. To verify such a connection, one may increase the dislocation mobility $B$ for the same stress rate ($\dot{\sigma}=E^{*}*10^{2}$/s). Figure~\ref{fig:3} (b) shows an enhanced drag effect (red curve), due to the increase of $B$ and a subsequent exponent change from 2.5 to 2.2. The drag effect may also be magnified when other dislocation mechanisms come into play, such as cross-slip. This can be seen in Fig.~\ref{fig:3} (b) blue curve, where a lower dislocation source density leads to a change of the $\tau$ exponent from 2.5 to 2.1.

\begin{figure}[t] \centering
	\includegraphics[width=\textwidth]{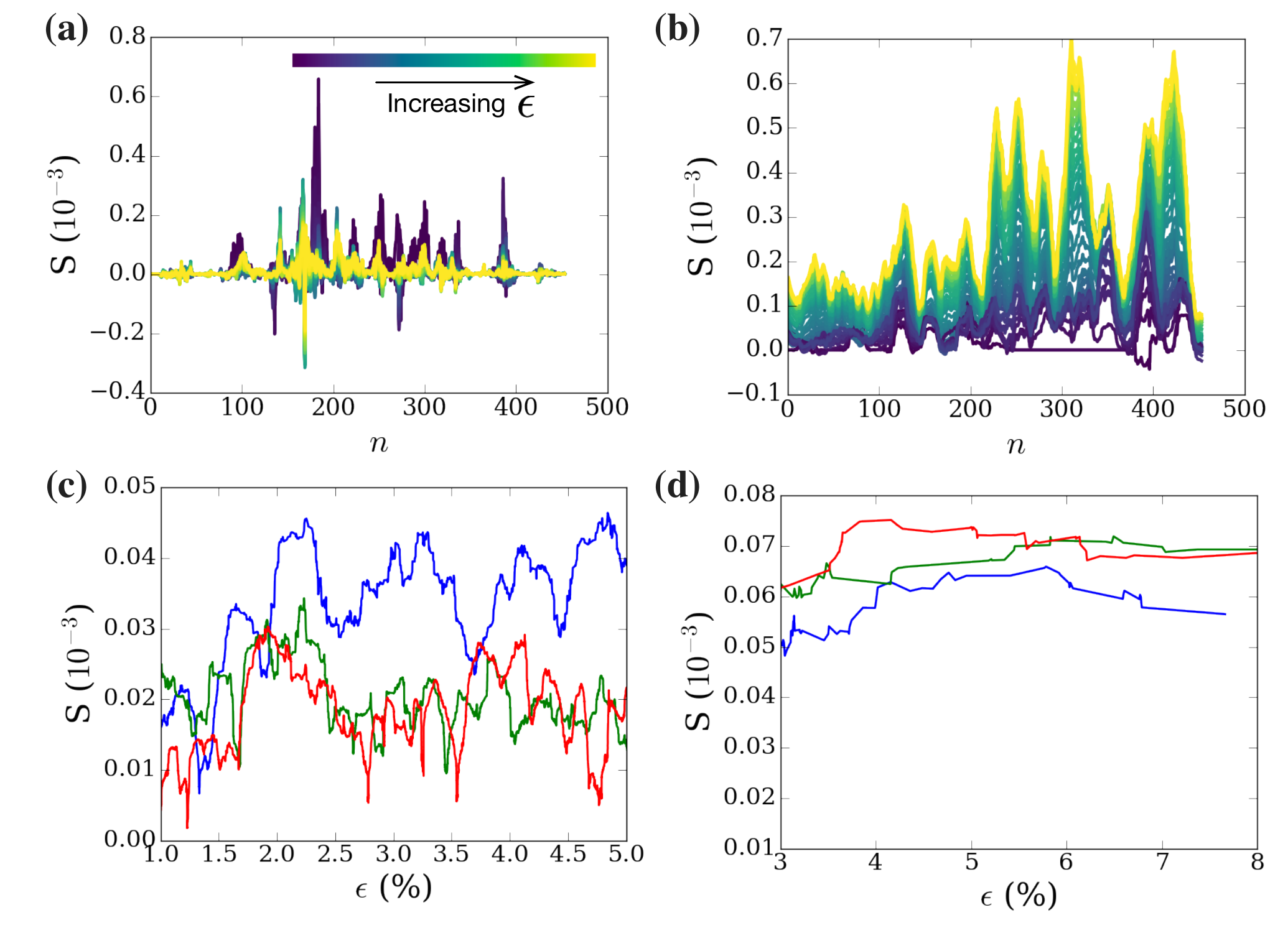}
	\caption{ {\bf Spatial and temporal event distribution in SC.}
		Event distribution on all slip planes during the loading up to 10\% strain for $\dot{\sigma}=E^{*}*10^{2}$ ({\bf (a)}) and for $\dot{\sigma}=E^{*}*10^{4}$ ({\bf (b)}).  The color changes from dark purple to yellow with increasing loading strain. {\bf (c)}: Average avalanche size for {  $\dot{\sigma}=E^{*}*10^{2}$} in a sample.
		{\bf (d)}: Average avalanche size for {  $\dot{\sigma}=E^{*}*10^{4}$} in a sample.}
	\label{fig:4}
\end{figure}

The aforementioned exponent crossover is associated to an onset of inhomogeneity along the boundaries of the finite volume, signifying spatio-temporally correlated plastic activity. At first sight, this is not unexpected since crystal plasticity is known to be unstable to strain localization, thus adding an inhomogeneity component to avalanche dynamics~\citep{asaro2006mechanics}. However, the combination of the exponent crossover with the onset of inhomogeneity in randomly evolving systems is uncommon. In Fig.~\ref{fig:4} (a) and (b), we show the spatial distribution of events along all slip planes $n$ for the loading process. Fig.~\ref{fig:4} (a) shows the event spatial distribution for loading rate of  $\dot{\sigma}=E^{*}*10^{2}$. Events are localized around certain slip planes, and furthermore, do not always happen at the same slip planes. For higher loading rate of $\dot{\sigma}=E^{*}*10^{4}$, the event distribution is more uniform among slip planes, shown in Fig.~\ref{fig:4} (b). The event size with increasing strain in Fig.~\ref{fig:4} (c) unveils an oscillatory-like behavior at small stress rate which disappears at higher stress rates.

The observed behavior is akin to a mean-field integrated behavior~\citep{papanikolaou2017avalanches}, labeled as the onset of an {\it avalanche oscillator}~\citep{Papanikolaou2012} as the strain-rate {\it decreases}. The novel terminology is required to distinguish typical integrated depinning behaviors taking place at large loading rates in various systems~\citep{fisher1998collective}. In this model, at low strain-rates, critical exponents $\tau$ and $\alpha$ are higher than mean-field, but the spectral density $x$~\citep{Papanikolaou2011} remains at the mean-field limit at low rates while  x' implies integrated mean-field behavior~\citep{Papanikolaou2011}. This novel behavior might explain  large exponents in crystal plasticity of small grains in polycrystals~\citep{leb18} or  crystalline pillar experiments~\citep{sparks2018shapes}. 

It is interesting to compare the statistical behavior of this model to mean-field plasticity avalanche behavior~\citep{uhl2015universal, spap16}, shown in Table~\ref{tab:ii}. In comparison, the  presented model has free nanoscale boundaries and a timescale competition between dislocation nucleation and drag: these are model characteristics that are not typically included in mean-field avalanche models.  Overall, it is found that these differences lead to an integrated behavior that is driven by quasi-periodic avalanche bursts~\citep{Papanikolaou2012}. 

Finally, it is worth noting that the model is limited to small deformations, does not include other possible three dimensional dislocation motions and does not include  boundary roughness stress effects or thermal effects on obstacles/sources~\citep{song2019discrete,pap19}.  
\begin{table}[!ht]
\Large
	\centering
	\caption{{ {\bf Universality and Exponents}. Basic mean-field avalanche exponents characterize power-law behaviors in avalanche sizes $P(S)\sim S^{-\tau}$, durations $P(T)\sim T^{-\alpha}$, spectral response $S(\omega)\sim\omega^{-x}$ and average size-duration relationship $\langle S\rangle\sim T^{x'}$.}}
	\begin{adjustbox}{max width=1.0\textwidth}
		\begin{tabular}{| c | c | c |}
			\hline
			Exponent & Mean-Field Theory & Avalanche Oscillator \\
			\hline\hline
			$\tau$ & 3/2 & {\small Rate-Dependent}$> 3/2$    \\ \hline
			$\alpha$ &  2 & {\small Rate-Dependent}$> 2$ \\ \hline
			$x$ & 2 & 2 \\ \hline
			$x'$ & 2 & 1 \\ \hline
		\end{tabular}
	\end{adjustbox}
	\label{tab:ii}
\end{table}

\section{Size effects in micro-crystal plasticity}

\begin{figure}[t]
	\centering
	\includegraphics[width=0.95\textwidth]{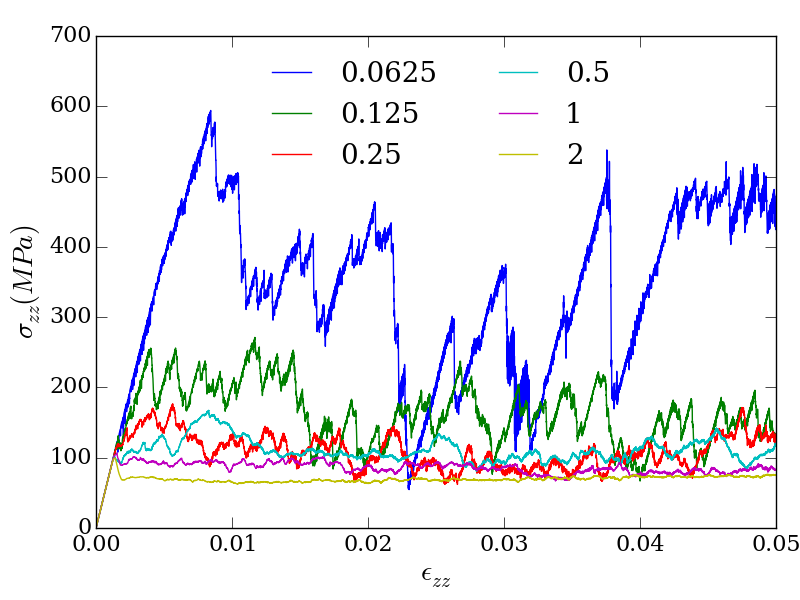}
	\caption{Axial stress--strain curves, $\sigma_{zz}$ vs $\epsilon_{zz}$. Strengthening and large stress drops emerge as $w$ decreases, with the width shown in the legend, in $\mu m$. }
	\label{fig:patterns}
\end{figure}

{   Experiments of uniaxial tension and compression in nanopillars have shown apparent material strengthening with decreasing pillar width $w$, with the yield strength varying as $\sigma_Y\sim w^{-n}$ with $n\in(0.4,0.8)$~\citep{Uchic:2009aa,Greer:2011aa}. The basic overall explanation behind size effects has been the gradual exhaustion of dislocation multiplication mechanisms, as the finite volume becomes smaller. A variety of possible mechanisms can explain most of the existing experimental phenomenology on strength size effects. However, the non-smooth post-yielding plasticity behaviors have been known to display size effects as well. Analysis of the statistics of abrupt plastic events has revealed that nanopillar events, appear to follow power-law distributions for strain steps with a large event cutoff that depends on specimen width~\citep{Weiss2000,miguel2001complexity,Miguel2001,weiss2003three}. These findings have evaded a unified model explanation until recently~\citep{Papanikolaou2017}.}

In this study~\citep{Papanikolaou2017}, 2D-DDD simulations of uniaxial compession for varying pillar widths $w$ ranged from $0.0625\mu m$ to $1\mu m$. In Figure~\ref{fig:patterns}(a) typical stress-strain curves are shown, with strengthening and large flow stress fluctuations as $w$ decreases. Due to strain-controlled loading in simulations, avalanches are captured as stress drops. 
{   The total number of observed avalanches is not controlled, however the total simulated strain is. Typically, these model simulations are performed up to 10 $\%$ strain for any dislocation density. As an example, for large dislocation densities~($\rho$~=~$10^{14}/{\rm m}^2$), a single sample volume may respond to uniaxial compression through {$10^3$} avalanches during strain-controlled loading.} 

As shown in Fig.~\ref{fig:size-effect}, the simulations identify clear size effects in the yield strength. Fig.~\ref{fig:size-effect}(a) shows that the yield strength $\sigma_Y$ decreases with increasing $w$. For pillar aspect ratio $\alpha=4$ (black line) we see a clear power-law dependence $\sigma_Y \sim w^{-0.45}$ which is similar to experimental observations. Morever, the sample strength depends on the aspect ratio $\alpha$, as also identified in experiments~\citep{senger2011aspect}. According to Fig.~\ref{fig:size-effect}(b), the yield strength decreases strongly with a power law $\sigma_Y\sim \alpha^{-0.36}$ for small widths, while in larger samples, this dependence is virtually absent. 
\begin{figure}[htb] 
	\centering
	\subfigure[]{
		\includegraphics[width=0.48\textwidth]{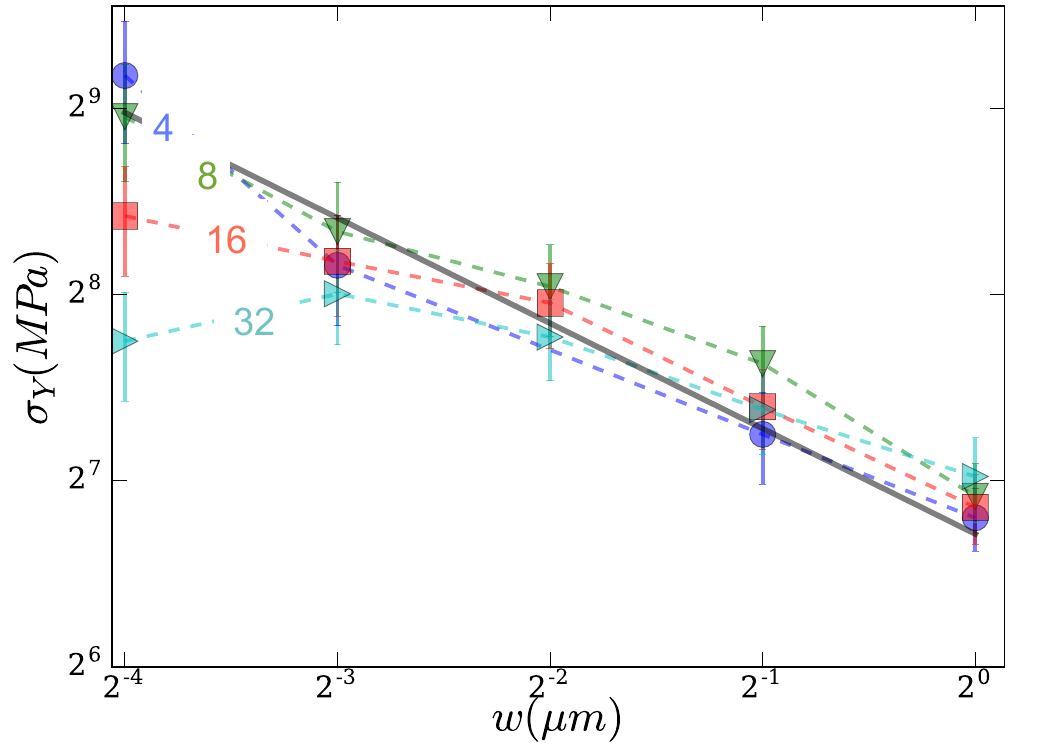}
	}
	\subfigure[]{
		\includegraphics[width=0.48\textwidth]{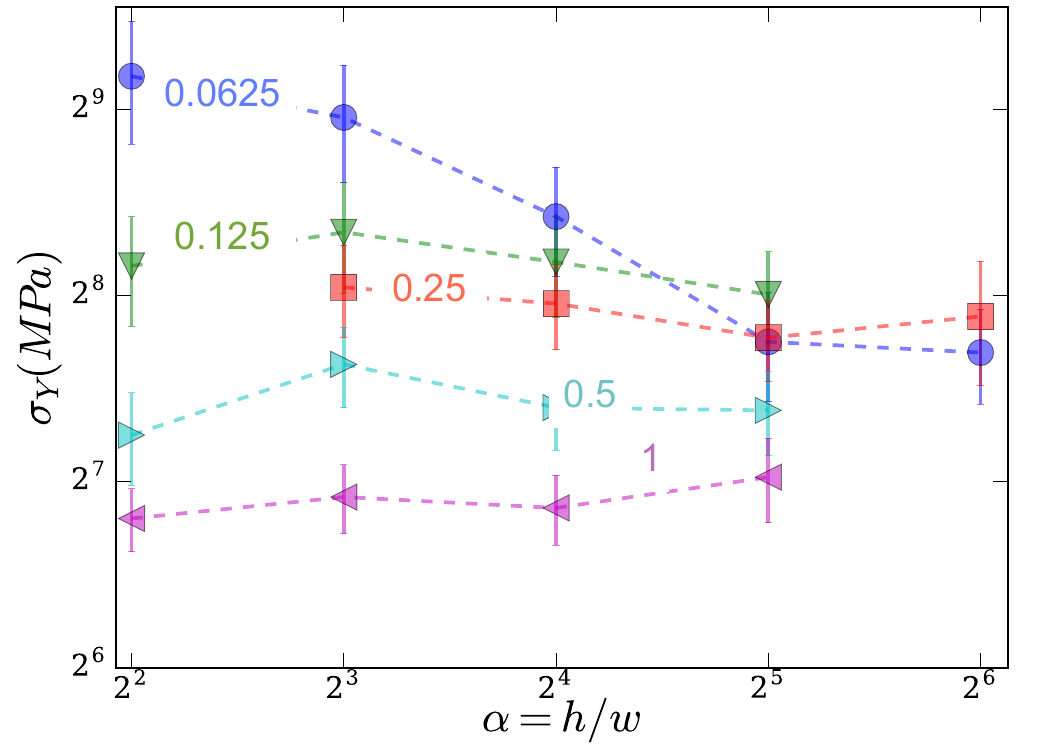}
	}
	\caption{(a) The width dependence of the yield stress. Different aspect ratios $\alpha$ are indicated by colored numbers. The fit to $\alpha=4$ is shown in bold black line, (b) the dependence of the yield stress on the aspect ratio $\alpha$.}
	\label{fig:size-effect}
\end{figure}

Avalanche behavior in the model of~\citep{Papanikolaou2017} is shown through power law tails of the event probability distributions $P(S)\sim S^{-\tau}{\cal P}(S/S_0)$. The onset of power-law behavior at decreasing $w$ is seen in Fig.~\ref{fig:statistics} (a) with an exponent $\tau=1.2\pm 0.2$ while $S_0 \sim w^{-1}$.  The existence of power-law behavior in the asymptotically small width limit becomes apparent in samples with low aspect ratio, as shown in the inset of Fig.~\ref{fig:statistics} (a), where the average event size $S_{av}\sim 1/w$ line is shown as a guide to the eye.
 
In Fig.~\ref{fig:statistics}(b), avalanche behavior statistics $P(S)$ are shown for three widths ($0.0625$, $0.25$ and $1\mu m$) and two aspect ratios ($4$ and $32$). Power-law behavior for varying aspect ratio is seen for the smallest system size $w=0.0625\mu m$; For larger systems, the distribution displays larger event sizes as the aspect ratio increases. This tendency is also seen in the behavior of $S_{av}$ (inset of Fig.~\ref{fig:statistics} (b)), where aspect ratio independence is observed for small widths ($0.0625$ and $0.125\mu m$), while for large widths there is a trend $S_{av}\sim \alpha^{1}$ (shown as a guide to the eye). 

\begin{figure}[t!]
	\centering
	\subfigure[]{
		\includegraphics[width=0.48\textwidth]{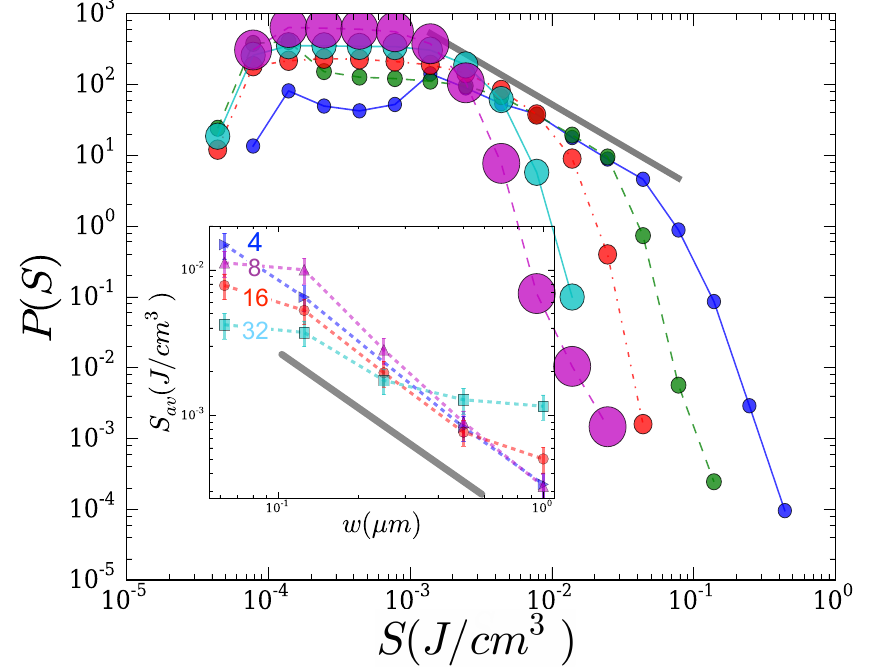}
	}
	\subfigure[]{
		\includegraphics[width=0.48\textwidth]{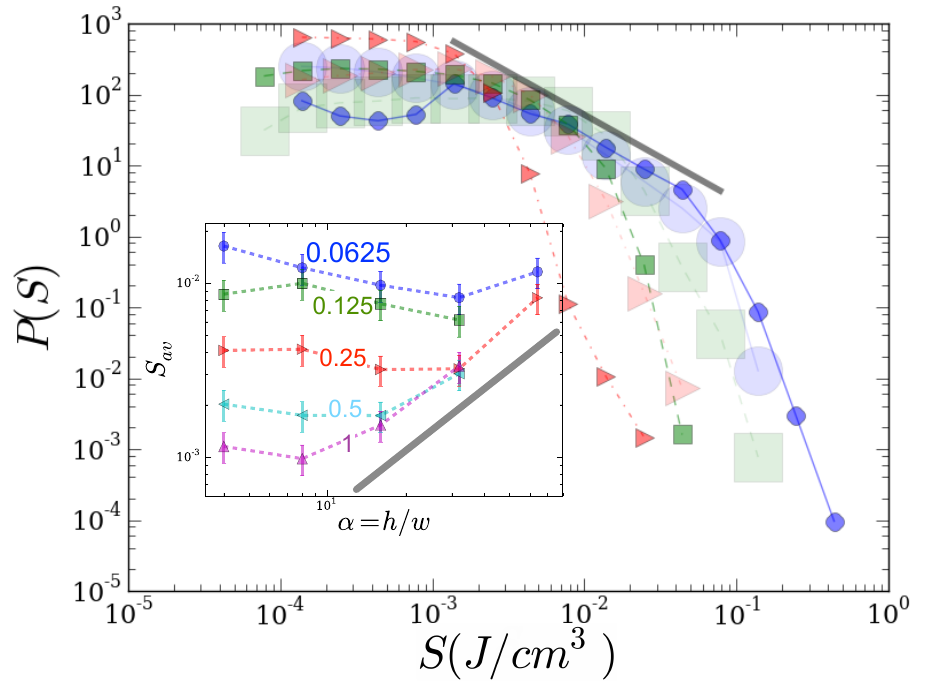}
	}
	\caption{
		{{\bf Histograms of abrupt events and cutoff dependence.}} (a) Width dependence of $P(S)$, demonstrating a power-law distribution as $w$ decreases for $\alpha=4$ (symbol size reflects $w$). In the inset, the average event size is shown as a function of $w$ for different aspect ratios $\alpha$. (b) Dependence of abrupt event statistics on pillar aspect ratio $\alpha$. Three different widths (
		$\bullet: w=0.0625\mu m$, 
		$\filledmedsquare: w=0.25\mu m$, 
		$\filledmedtriangleright:w=1\mu m$) are shown for two different aspect ratios $\alpha=4$ and $32$ (the symbol sizes follow the aspect ratio's magnitude for clarity).}
	\label{fig:statistics}
\end{figure}

\section{Unveiling the crystalline prior deformation history using unsupervised machine learning approaches}
\label{sec:unsupervised}
{   Elements of prior deformation history in crystals are needed for any prediction of mechanical properties in plasticity. The most common example is the accumulated dislocation density, which is typically used for the prediction of flow stress. It is natural to expect that a wealth of additional mechanical property predictions can be made through the use of multi-dimensional deformation information, possibly originating in {\it in-situ} strain maps. However, the efficient and systematic development of such mechanical property predictions requires data-intensive dimensional reduction and classification that has been common in machine learning (ML) methods.}

ML methods have been recently used in science and engineering~\citep{decost2017computer,ramprasad2017machine,mueller2016machine,pilania2013accelerating} and may predict microstructural properties~\citep{pilania2013accelerating}, optimize material design~\citep{liu2015predictive} and infer deformation history~\citep{pap19}. The usage of ML in mechanical deformation studies started from analyzing nanoindentation responses towards the prediction of material properties~\citep{khosravani2017development,iskakov2018application,meng2015identification,meng2017objective,huhn2017revealing}. In a new direction on this topic, a recent work~\citep{pap19} showed that the analysis of {\it small}-deformation strain correlation images may unveil the prior deformation history of materials. This process, which can be built on any version of DIC~\citep{Schreier:2009qy}, shows that the use of unsupervised ML methods on strain correlations, may  establish an equation free approach for the recognition of prior deformation history for large sample width $w$. The reason for the method's effectiveness is the fact that the primary features of crystal plasticity, such as spatial strain gradients in the microstructure, may also be reflected in spatially resolved strain correlations~\citep{chaikin1995principles,papanikolaou2013isostaticity,papanikolaou2007quantum,raman2008quantum}. {   In three dimensions, it is expected that multiple cross-sections' strain information would be required for analogous method effectiveness.}

The method was implemented~\citep{pap19} in an explicit model of 2D-DDD, where two slip systems are used, for $50$ random initializations of sources and obstacles and 0.1~\%, 1~\% and 10~\% {\it prior} loading of the samples. In this way, statistically reliable initial conditions are produced at various initial dislocation densities. 
\begin{figure}[t] \centering
	\includegraphics[width=\textwidth]{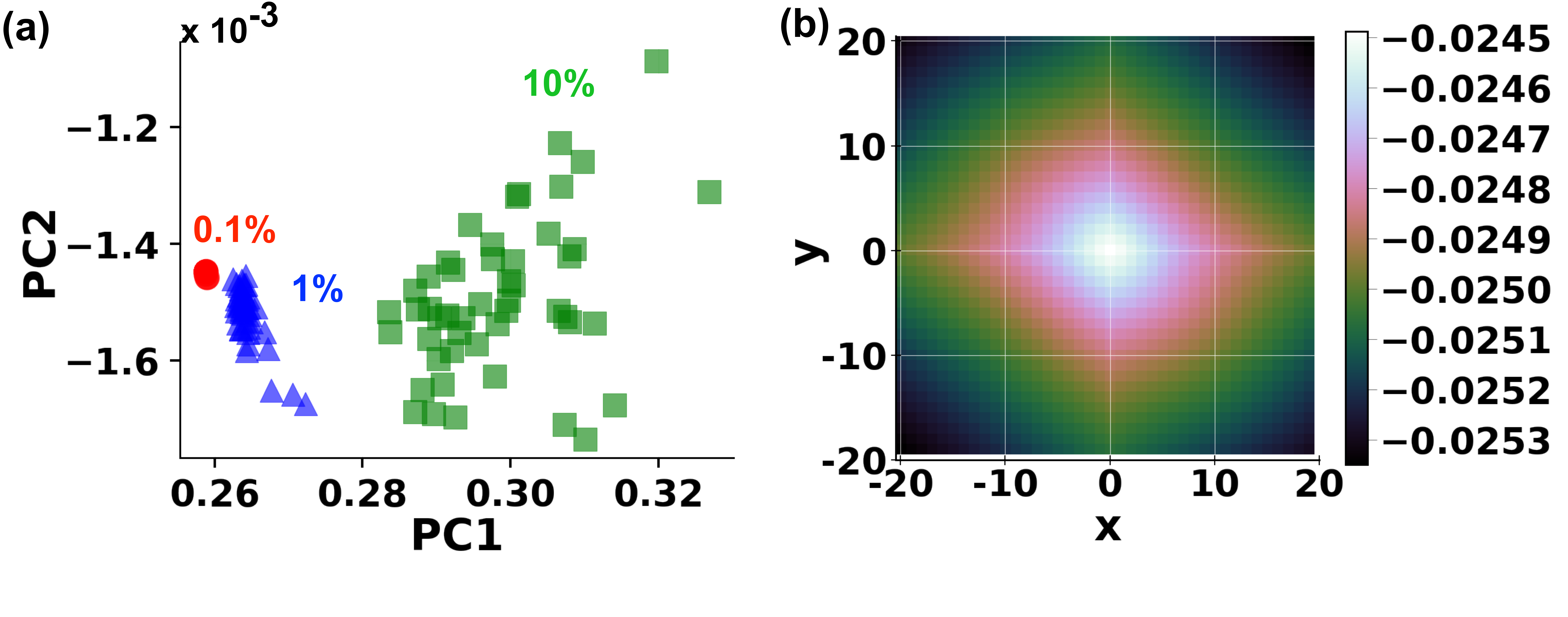}
	\caption{\textbf{$\boldsymbol{w=1} \boldsymbol{\mu m}$ -- Prior deformation history of samples (see~\citep{pap19}):} (a) Red $\bullet$ are samples with $0.1~\%$ prior strain, blue $\filledmedtriangleup$ samples with $1~\%$ prior strain and green $\filledmedsquare$ are samples with $10~\%$ prior strain. (b) First principal component of PCA, shown in sample coordinates (Fig.~\ref{fig:schematic} (b)). Colormap is unitless.}
	\label{fig:width1}
\end{figure}
\begin{figure}[h] \centering
	\includegraphics[width=\textwidth]{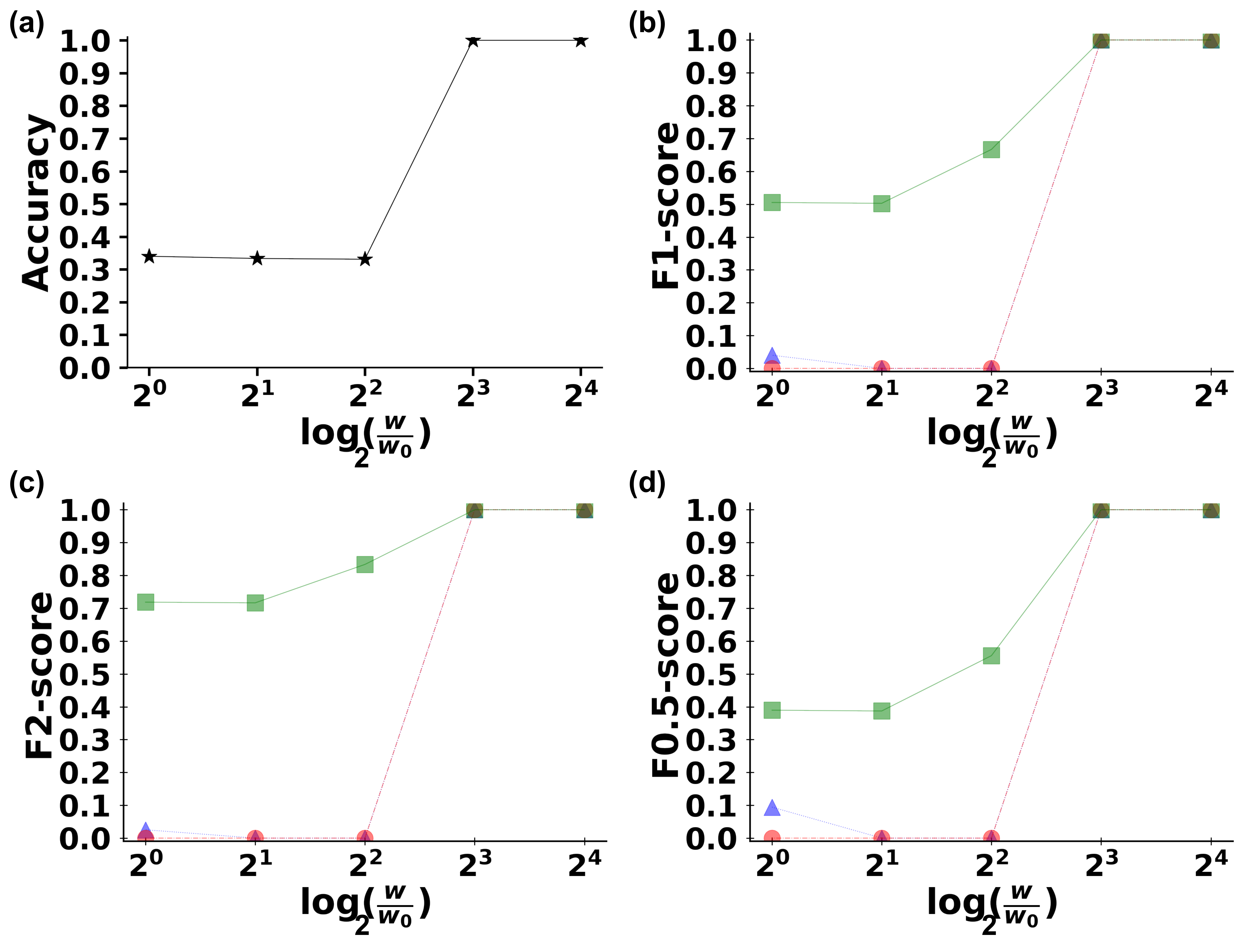}
	\caption{{\textbf{Measures of success for classification of samples (see~\citep{pap19}):}} (a) Accuracy score for the samples. (b) $F_1$-score of the 3 clusters that are formed. (c) $F_2$-score of the 3 clusters that have formed. (d) $F_{0.5}$-score of the 3 clusters. Red $\bullet: \epsilon_{prior} = \ 0.1 \%$, blue $\filledmedtriangleup: \epsilon_{prior} = \ 1 \%$ and green $\filledmedsquare: \epsilon_{prior} = \ 10 \%$ }
	\label{fig:accuracy}
\end{figure}
The prior-deformed samples are subjected to a small compressive but {\it non-invasive} load of 0.1~\% testing strain, and the final strain images consist of the {\rm tests}. The applied strain is small so that it does not introduce significant further plastic deformation on the samples . After removing the strain information present at the unload stage, corresponding to well-annealed samples, strain correlation signatures are examined on strain profiles (see~ Fig.~\ref{fig:schematic}~(b)) created by the small load mechanical testing and are collected in a matrix $\matr D$:
\begin{equation} 
\centering
\matr D =
\begin{bmatrix}
C^{[1]} [r_1|h_ih_j] \cdots C^{[1]} [r_m|h_ih_j]
\\
\vdots
\\
C^{[n]} [r_1|h_ih_j] \cdots C^{[n]} [r_m|h_ih_j] 
\end{bmatrix}
\end{equation}
where each row of $\matr D$ contain the vector $\matr d_{ij}$:
\begin{equation}
\matr d_{ij} = (C^{[k]}[r_1|h_ih_j],  C^{[k]}[r_2|h_ih_j], \cdots,  C^{[k]}[r_m|h_ih_j])
\end{equation}
where the correlation function $C^{[k]}[r_v|h_ih_j], k=1,n $ ($n \ = $ No. of samples) is modeled after the Materials Knowledge System in Python (PyMKS~\citep{Wheeler:2014aa}) scheme. '

{   Due to the inherent 2D nature of the problem presented in~\citep{pap19},  shear band formation upon small reloading may be picked up in large samples using spatial correlations, for all prior deformation levels. Similar ML schemes may be used in a 2D cross-section of 3D problems, for example on surface deformation fields of nanoindented samples. However, in 3D settings, a similar ML scheme would require the study of multiple volume cross-sections, in order to characterize prior deformation, or other mechanical properties. }

The validity of the ML workflow is quantified through the investigation of accuracy and $F_{\beta}$ scores~\citep{Baeza-Yates:2011aa}. {\it Accuracy} is defined as the fraction of correct predictions of the classifier. The $F_{\beta}$ scores are used to quantify the performance in each cluster:
\begin{equation}
F_{\beta} = (1+ \beta^2)\cdot\frac{p\cdot r}{(\beta^2 \cdot p) + r}
\end{equation}
where precision $p$ is the number of correctly classified samples in a cluster divided by the number of all classified samples in the cluster, and recall, $r$, is the number of correctly classified samples in a cluster divided by the number of samples that should have been in that cluster.

Results for larger systems are shown in Fig.\ref{fig:width1}, where we observe the results of unsupervised ML for systems sizes of $w = 1 \ \mu$m (Fig.~\ref{fig:width1} (a)) and the corresponding smooth correlations (Fig.~\ref{fig:width1} (b)). The unsupervised ML results for all system sizes can be summarized in Fig.~\ref{fig:accuracy}, where the accuracy and $F_{\beta}$ scores are shown. Maximum value 1 means that all samples have been correctly classified.  For the $F_2$-score, weight of $r$ is increased, and the 0.7 maximum value is expected for the ``square'' cluster of smaller system sizes. For the $F_{0.5}$-score the weight of $r$ is decreased. A correspondence between strain correlations and prior deformation history is found with 100 \% success for large systems, .

\section{Predicting the mechanical response of crystalline materials using supervised machine learning}
\label{sec:supervised}
{   While unsupervised ML is necessary when the number of distinct data classes is unknown, supervised ML can perform much improved classification tasks.
 In this section, we discuss the application of supervised ML approaches on the dataset of~\citep{pap19}, assuming known prior deformation histories of 80\% of the samples. The aim is to identify relationships that fully describe the connections between strain correlations and prior processing history. 
 
 In addition, the understanding and classification of prior deformation is equivalent to knowing the deformation {\it State}. In that case, one should be able to perform predictions of future mechanical response, albeit at average levels. We show that we can {\it statistically} predict mechanical responses for test data (20\% of the samples), which can be thought of as average future mechanical responses of  classified specimens. }
\begin{figure}[h] 
\centering
	\includegraphics[width=\textwidth]{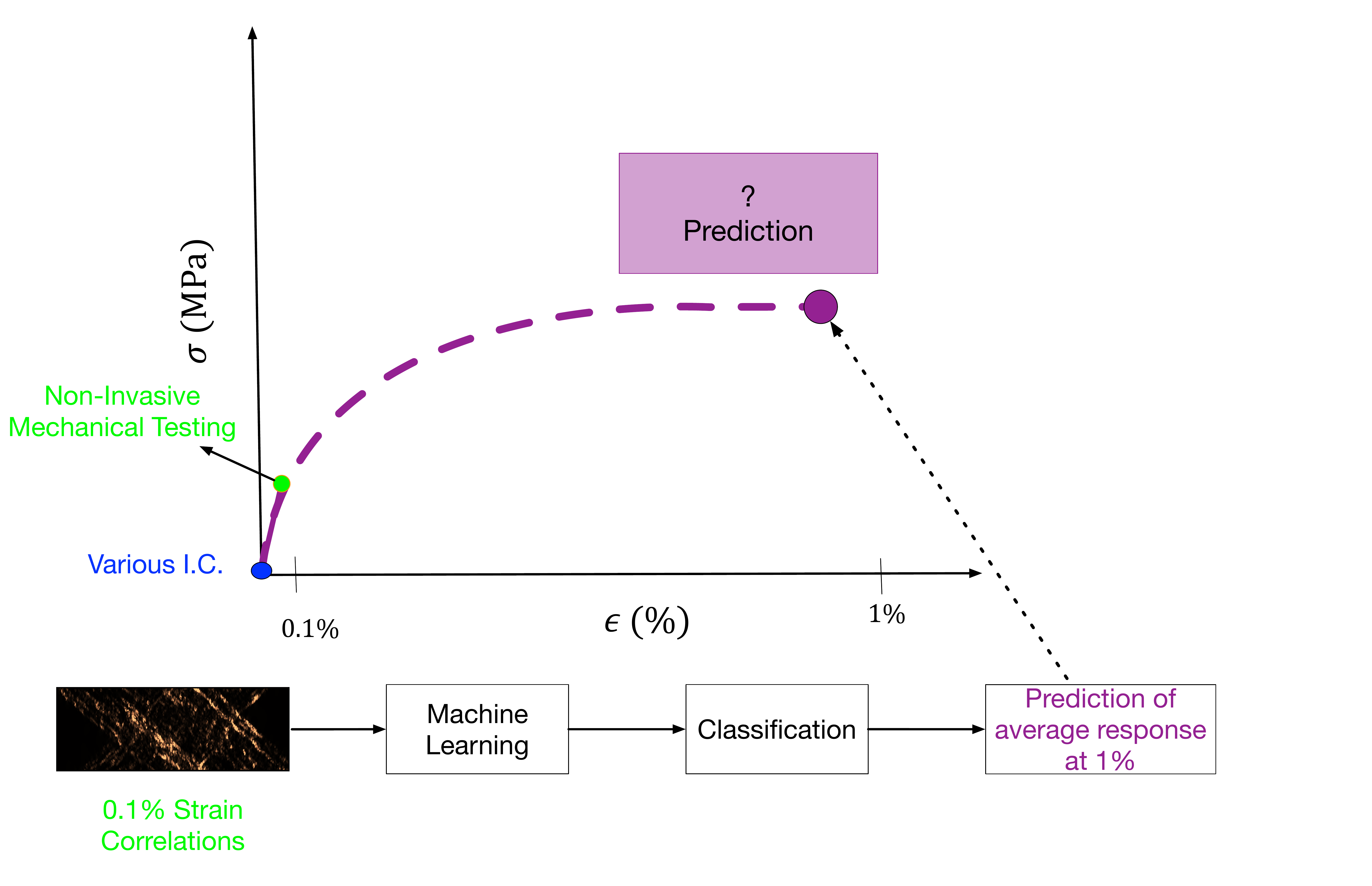}
	\caption{\textbf{Schematic for obtaining the 1\% strain mechanical response of unknown samples}}
	\label{fig:schematic_prediction}
\end{figure}

In supervised ML approaches, the dataset consists of samples with known outputs/features (in our case, prior deformation history is known for each sample), and the goal is to create robust algorithms that recognize sample/feature correspondences with high accuracy. In a typical supervised ML workflow, collected datasets are split to training and testing data sets. The algorithm is trained on the training data sets, and then it is tested as to the validity and accuracy of the testing data sets. In the absence of big data collections, it is common to perform an 80~\%~-20~\%-split for the training and testing data sets.

We train two types of supervised ML algorithms on the training set: Neural networks~\citep{bishop1995neural} and Decision Trees~\citep{quinlan1986induction}. Neural Networks are a set of algorithms, modeled loosely after the human brain, that are designed to recognize patterns in datasets and consist of ``neurons" from which the dataset passes through and activates various input functions. Decision trees are a set of decisions for the features of the input matrix, modeled after trees. The algorithm finds patterns in the features and creates leaves of a tree. When all possible patterns have been found, we have multiple leaves in a tree, hence the name decision tree.

The most accurate Neural Networks and Decision Trees  may be found through a parameter search using an algorithm for parameter optimization (GridSearchCV~\citep{Bergstra2011}). The GridSearchCV algorithm allows the input of multiple parameters of a given classifier, and output of the set of parameters that will provide the highest accuracy for the problem.
\begin{table}[t]
	\Large
	\centering
	\caption{{ {\bf Neural Networks}. Accuracy scores on supervised machine learning for identification of prior deformation histories via spatial strain correlations.}}
	\begin{adjustbox}{max width=1.0\textwidth}
		\begin{tabular}{| c | c | c |}
			\hline
			\textbf{$w$} & \textbf{Training Set Accuracy} & \textbf{Test Set Accuracy} \\ \hline \hline
			\textbf{0.125 $\mu m$} & 90.38 \% & 83.3 \% \\ \hline
			\textbf{0.25 $\mu m$} & 91.35 \% & 100 \% \\ \hline
			\textbf{0.5 $\mu m$} & 100 \% & 100 \% \\ \hline
			\textbf{1 $\mu m$} & 100 \% & 100 \% \\ \hline
			\textbf{2 $\mu m$} & 100 \% & 100 \% \\ \hline
		\end{tabular}
	\end{adjustbox}
	\label{tab:iii}
\end{table}

We employed the use of the GridSearchCV~\citep{Bergstra2011} algorithm for Neural Networks~\citep{bishop1995neural} and Decision Trees~\citep{quinlan1986induction}, in order to identify the parameters that will produce the highest accuracy in the supervised problem. In the case of Neural Networks, the parameter search included adaptive or constant learning rate ranging from $10^{-5}$  to $10^{3}$. For Decision Trees, the input parameters  on the GridSearchCV algorithm were gini or entropy criteria with the maximum depth of the tree ranging from 12 to 16. With these parameters the highest accuracy was provided for adaptive learning rate of $10^{-5}$ for Neural Networks, while for Decision Trees, the best criterion was gini~\citep{quinlan1986induction} with maximum depth (for w = 0.125 $\mu m$) set at 14 leaves.

\begin{table}[h!]
	\Large
	\centering
	\caption{{{\bf Decision Trees}.Accuracy scores on supervised machine learning for identification of prior deformation histories via spatial strain correlations.}}
	\begin{adjustbox}{max width=1.0\textwidth}
		\begin{tabular}{| c | c | c |}
			\hline
			\textbf{$w$} & \textbf{Training Set Accuracy} & \textbf{Test Set Accuracy} \\ \hline \hline
			\textbf{0.125 $\mu m$} & 100 \% & 83.3 \% \\ \hline
			\textbf{0.25 $\mu m$} & 100 \% & 100 \% \\ \hline
			\textbf{0.5 $\mu m$} & 100 \% & 100 \% \\ \hline
			\textbf{1 $\mu m$} & 100 \% & 100 \% \\ \hline
			\textbf{2 $\mu m$} & 100 \% & 100 \% \\ \hline
		\end{tabular}
	\end{adjustbox}
	\label{tab:iv}
\end{table}

The scores for the supervised problem (See Tables~\ref{tab:iii},~\ref{tab:iv}), exceed the scores of the unsupervised problem reported on~\citep{pap19} (also see Fig.~\ref{fig:accuracy}). This result was expected since the deformation histories are now known for the training set, and it is easier to establish connections between known input-outputs. 

\begin{figure}[h] \centering
	\includegraphics[width=\textwidth]{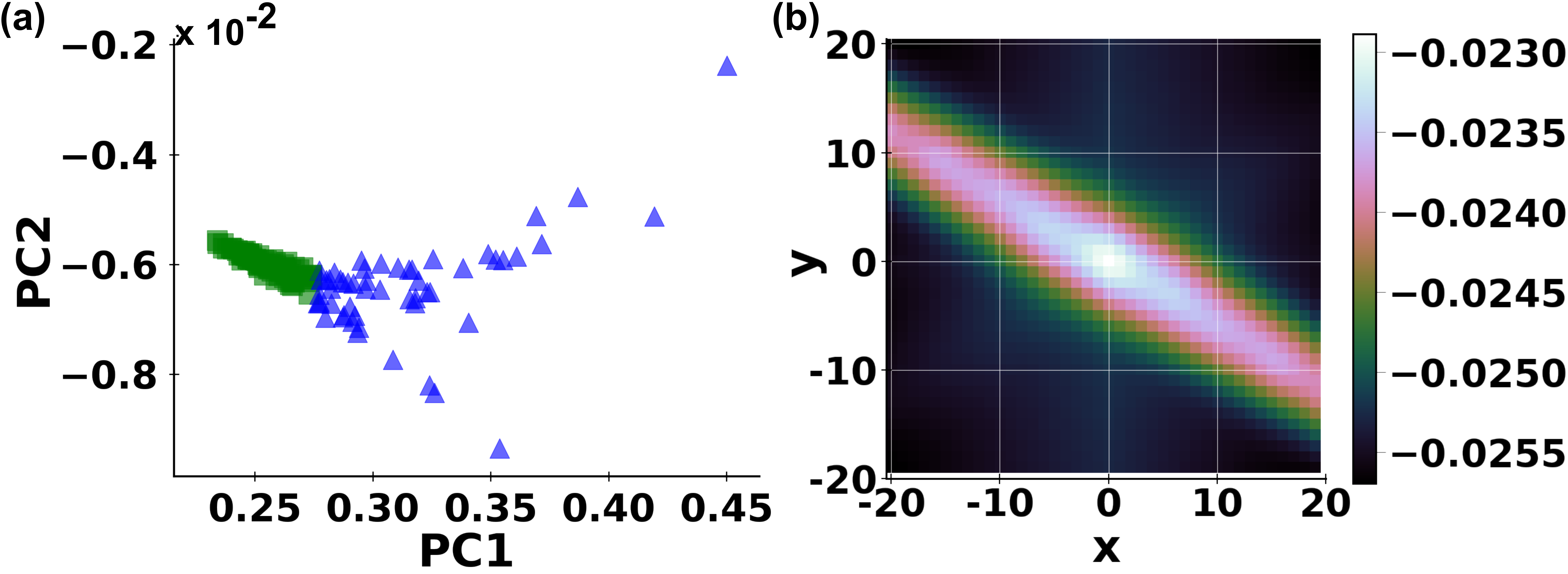}
	\caption{\textbf{$\boldsymbol{w=0.5} \boldsymbol{\mu m}$ - Prior deformation history of samples, large reload strain (1 \%) (see~\citep{pap19}):} (a) Colors follow Fig.~\ref{fig:width1}. The failure of the classifier is evident. (b) First principal component of PCA, shown in sample coordinates (Fig.~\ref{fig:schematic}). The anisotropy of the component is largely due to the high localization effects upon reloading to higher strain.}
	\label{fig:large_strain}
\end{figure}

With the application of supervised algorithms on the dataset, we are able to find a relationship between the known prior deformation histories (3 classes of uniaxial compressive strain) and spatial strain correlations in training samples and use it for the classification of testing samples with high accuracy. We assume that  samples that belong in each class are ``similar" in terms of their mechanical properties. We use classified samples as averages for the prediction of the mechanical response upon further compression. In Fig.~\ref{fig:schematic_prediction}, a schematic for the prediction of the mechanical response is shown, and we discuss the detailed process of calculating the average response based on prior deformation.

\begin{figure}[!ht] \centering
	\includegraphics[width=\textwidth]{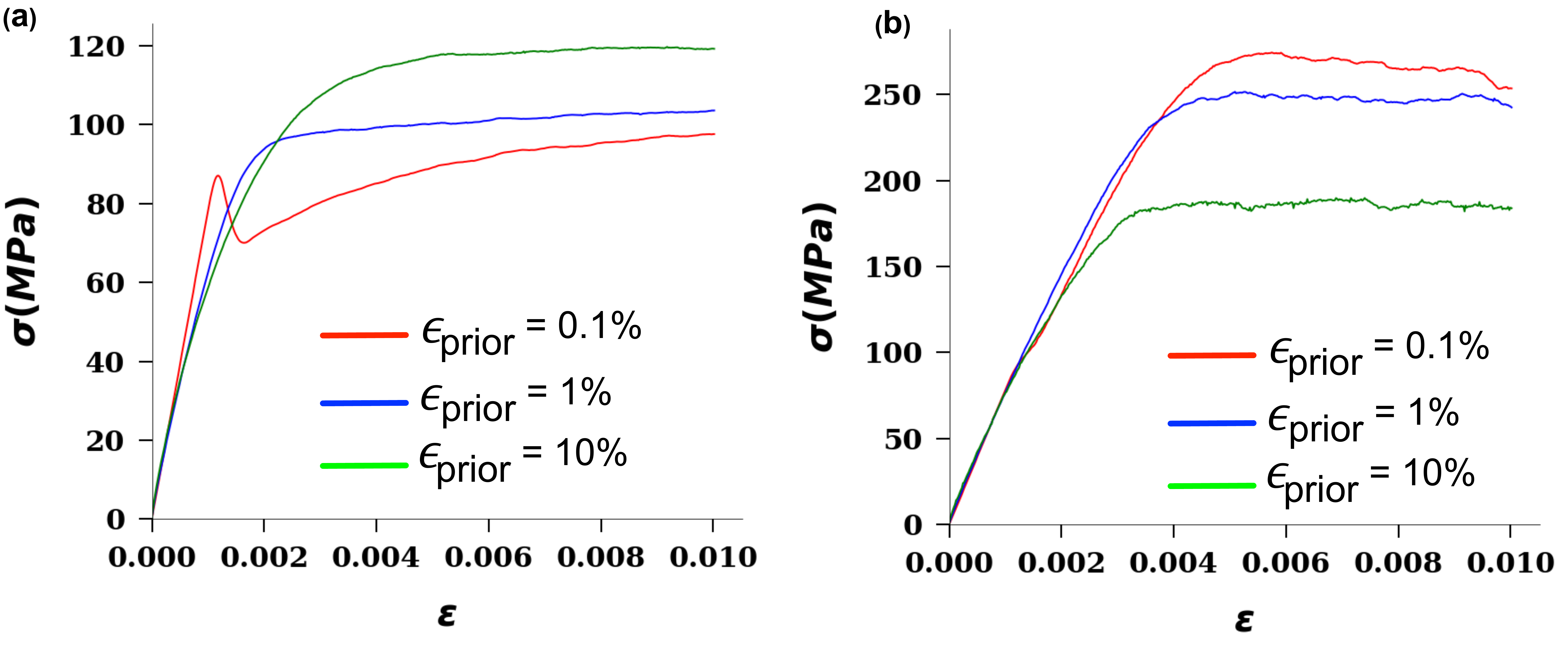}
	\caption{\textbf{Prediction of the mechanical response of samples with a known prior deformation history:} (a) $w = 2 \ \mu m$. (b) $w = 0.25 \ \mu m$.}
	\label{fig:prediction_curves}
\end{figure}

\begin{figure}[!ht] \centering
	\includegraphics[width=0.8\textwidth]{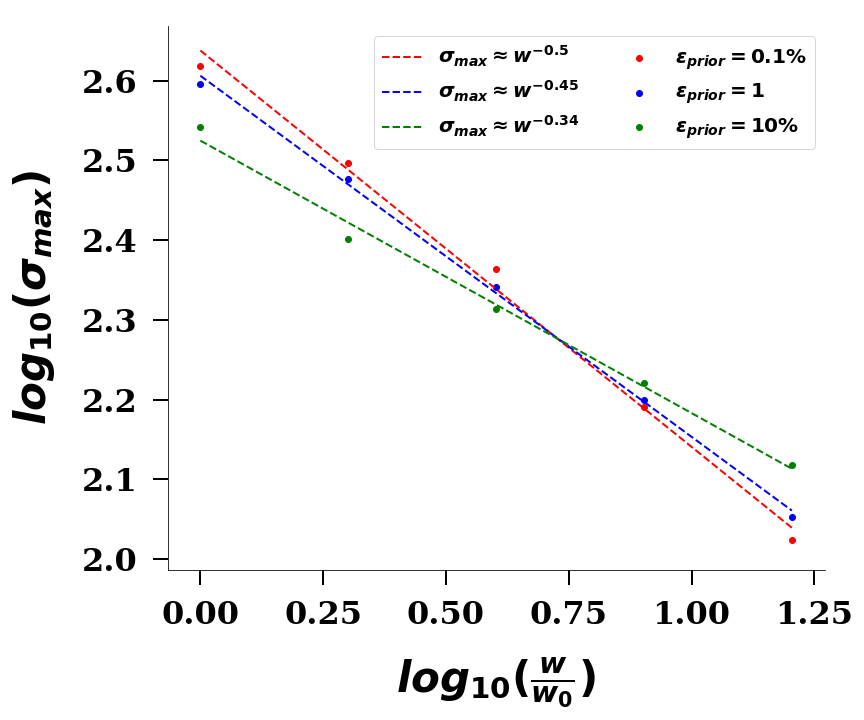}
	\caption{\textbf{Size effects in thin films:} The maximum predicted stress (see Fig.~\ref{fig:prediction_curves}) is plotted against sample widths. The relationships derived correspond to a power law with an exponent that changes depending on the degree of prior deformation history of specimens.}
	\label{fig:s-w-rel}
\end{figure}

We create 3 separate datasets, one for each deformation class (irrespective of the accuracy of the algorithm). For samples in each class, we assume future deformation features (1\% testing deformation) as known, since there is a one-to-one correspondence between testing deformation levels~\footnote{the same sample that is loaded to 0.1\% testing deformation to capture the strain correlation patterns is also loaded to 1\% testing deformation see~\citep{pap19}}. While the classification of datasets with 1\% reloading strain has not taken place (also see Fig.~\ref{fig:large_strain}), this is irrelevant to promoting predictions, since samples share the same initial dislocation ensemble, which may be found for small reload strain.

For each dataset, we collect the average reload response (1 \% strain) per width. The results can be seen in Figs.~\ref{fig:prediction_curves},~\ref{fig:s-w-rel}. In Figure~\ref{fig:prediction_curves}, we observe, for decreasing width, whether the prior deformation history controls the mechanical response and the hardening behavior of the material. Red lines are for $\epsilon_{prior} = \ 0.1\%$, blue lines for $\epsilon_{prior} = \ 1\%$ and green lines for $\epsilon_{prior} = \ 10\%$. We see a transition at $w \simeq 1 \ \mu m$, where the maximum stress response in further deformation, changes from being high deformation history dominated (prior deformation history = 10\%, Fig.~\ref{fig:prediction_curves} (a)) to low deformation history dominated (prior deformation history = 0.1\%, Fig.~\ref{fig:prediction_curves} (b)).

Prior deformation history of samples directly connects to the relaxed dislocation configuration in the volume upon unloading, with prior-strain levels indicating corresponding dislocation density levels in the crystal. 
In Fig.~\ref{fig:s-w-rel}, the sample yield stress is plotted against the thin film width for different dislocation density levels (acquired through prior deformation), which demonstrate an {\it evolving} size effect  $\sigma \approx w^{-a}$, where $a$ is shown in the legend, with $a\rw0$ as the dislocation density increases. This is consistent with the basic phenomenological expectation in crystalline size effects' literature~\citep{El-Awady2015,papanikolaou2017avalanches,PhysRevLett.122.178001}.  It is worth noting that the discussed model in this work is the first discrete dislocation model demonstration of this well suggested transition (since Taylor) as a function of pre-existing dislocation density. The origin and further consequences of these findings will be discussed elsewhere.

Overall, the suggested approach for the prediction of  mechanical responses implies there is an accurate method to describe and predict far-from-equilibrium mechanical-response phenomena: Given a sample of unknown origin, and a known database of prior deformation histories, one only needs to apply a small load mechanical test, capture spatial strain correlation features, use them as part of the test set in the supervised ML problem and obtain a prediction of future mechanical response, and the prior deformation history/dislocation density of the crystal.

\section{Summary}
\label{sec:summary}
In this chapter, we presented recent advances in the multiscale modeling of material science to understand how and when crystal plasticity of small finite volumes, displays dependence on loading rate, specimen size and pre-existing, load-induced, dislocation microstructures. We introduced and discussed an explicit model of discrete dislocations which is both minimal (in model details) and rich (in results and conclusions). While we investigated only the simple example of uniaxial compression, the model is directly generalizable to any other geometry in mechanics. Intrinsic, plasticity-induced crackling noise allows for thorough, statistically reliable examination of event statistics in a finite-volume system. Through an extensive investigation of this model, there has been a thorough and deep understanding of the collective effects in nanocrystal plasticity. The ultimate results of these studies have been the development of predictions for further signatures of rate and size effects, especially the finding of a dislocation-density dependent size-effect that promotes a transition to Taylor work hardening for the very first time in discrete dislocation modeling efforts. In addition, a major result of these studies has been a precise machine-learning method for mechanical predictions of deformation characteristics. Nevertheless, beyond particular predictions in one or another aspect of nanocrystal plasticity, the most important accomplishment of these research efforts has been the unified investigation of small finite-volume nanocrystal plasticity as a whole, through rates, sizes and prior deformation histories.

\newpage
\bibliographystyle{agsm}

\end{document}